\documentclass{article}
\usepackage{authblk}
\def\bd{\begin{displaymath}}\def\ed{\end{displaymath}}
\def\be{\begin{equation}}\def\ee{\end{equation}}
\def\bea{\begin{eqnarray}}\def\eea{\end{eqnarray}}
\def\ba{\begin{array}}\def\ea{\end{array}}
\def\bs{\begin{split}}\def\es{\end{split}}
\def\lb{\label}

\def\a{\alpha}\def\b{\beta}\def\d{\delta}
\def\f{\phi}\def\g{\gamma}
\def\k{\kappa}\def\l{\lambda}\def\m{\mu}\def\n{\nu}\def\r{\rho}\def\s{\sigma}
\def\y{\eta}

\def\D{\Delta}
\def\S{\Sigma}

\def\lra{\leftrightarrow}\def\bdot{\!\cdot\!}\def\ha{{1\over 2}}
\def\coo{coordinates }\def\cc{coupling constant }\def\ie{i.e.\ }
\def\poi{Poincar\'e }\def\des{de Sitter }

\def\PL#1{Phys.\ Lett.\ {\bf#1}}\def\CMP#1{Commun.\ Math.\ Phys.\ {\bf#1}}

\def\PR#1{Phys.\ Rev.\ {\bf#1}}\def\CQG#1{Class.\ Quantum Grav.\ {\bf#1}}

\def\JMP#1{J.\ Math.\ Phys.\ {\bf#1}}

 \def\IJMP#1{Int.\ J. Mod.\ Phys.\ {\bf #1}}

\def\JHEP#1{JHEP\ {\bf#1}}
\def\RMP#1{Rev.\ Mod.\ Phys.\ {\bf#1}}
\def\arx#1{{\tt arXiv:#1}}

\def\LRR#1{Liv.\ Rev.\ Rel.\ {\bf#1}}

\def\cD{{\cal D}}
\def\hx{\hat x}\def\hp{\hat p}
\def\xp{\,x{\cdot}p\,}\def\px{\,p{\cdot}x\,}
\def\bt{{\bar t}}\def\bs{{\bar s}}\def\mn{{\mu\nu}}

\def\bdot{\!\cdot\!}

\def\cD{{\cal D}}

\def\hx{\hat x}\def\hp{\hat p}

\def\xp{\,x{\cdot}p\,}\def\px{\,p{\cdot}x\,}

\def\bt{{\bar t}}\def\bs{{\bar s}}

\begin{document}
\title{Yang model revisited}
\vskip80pt
\author[1,2]{S. Mignemi{\footnote{smignemi@unica.it}}}
\affil[1]{Dipartimento di Matematica e Informatica, Universit\`{a} di Cagliari, \newline via Ospedale 72, 09124 Cagliari, Italy}
\affil[2]{INFN, Sezione di Cagliari, Cittadella Universitaria, 09042 Monserrato, Italy}
\maketitle

\begin{abstract}
A long time ago C.N. Yang proposed a generalization of the Snyder model to the case of a curved background spacetime,
based on an algebra isomorphic to $so(1,5)$, which includes as subalgebras both the Snyder and the de Sitter algebras.
His proposal can therefore be interpreted as a model of noncommutative curved spacetime, and could be useful for
relating physics at very small and very large scales.

We review this model and some recent progress concerning its generalizations and its interpretation in the
framework of Hopf algebras. We also report some possibilities to relate it to more phenomenological aspects.

\end{abstract}
\vskip60pt

\section{Quantum gravity and noncommutative geometry}
At present, no complete theory of quantum gravity is available.
 However, it is known that the predictions of quantum mechanics and general relativity imply the existence
of a minimal measurable length of the size of the Planck length $L_P=\sqrt{\hbar G\over c^3}=10^{-33}$ cm \cite{gar}.
Hence it is natural to believe that the properties of spacetime at this scale must be rather different from those
we observe in everyday experience.

Among the proposals for a model of spacetime at Planck scale, noncommutative geometry plays a relevant role \cite{nonc}.
Noncommutative geometry is based on the assumption that the components of  the position operator do not commute,
leading to the impossibility of  localizing a particle exactly.
Among various approaches to this field, a relevant one is the Hopf algebra formalism \cite{hop}, which is suitable
for the description of the symmetries of the geometry.

Noncommutative geometries are usually defined on flat spacetime, but their extension to
curved spacetime has earned some interest recently because of possible implications for astrophysical observations,
like time delay of photons from distant sources \cite{ros}.
However, also the formal aspects of this extension are noticeable, in particular the relations between the curvature
of spacetime and of momentum space.
Moreover, these models relate the properties of spacetime at microscopic and macroscopic scales.

The first model of this kind was proposed by C.N. Yang already in 1947 \cite{yan}.
In the present paper, we review this framework and discuss some recent progresses and generalizations.

\section{The Snyder algebra}
We start by summarizing the main features of the Snyder model, from which Yang took inspiration.
In 1947 Snyder proposed the first model of noncommutative geometry \cite{sny}.
His aim was to construct a theory that included a fundamental length without breaking the Lorentz invariance.
This purpose was realized by deforming the commutation relations of the Heisenberg algebra.

In fact, the model was defined through an algebra that, besides the deformed Heisenbeg algebra generated by positions
$\hx_\m$ and momenta $\hp_\m$, contained the Lorentz algebra with generators $J_\mn$,
\bd
[\hx_\m,\hx_\n]=i\b J_\mn,\quad[\hp_\m,\hp_\n]=0,\quad[\hx_\m,\hp_\m]=i(\y_\mn+\b\hp_\m\hp_\n),
\ed
\bd
[J_\mn,J_{\r\s}]=i\big(\y_{\m\r}J_{\n\s}-\y_{\m\s}J_{\n\r}+\y_{\n\r}J_{\m\s}-\y_{\n\s}J_{\m\r}\big),
\ed
\be\lb{1}
[J_\mn,\hp_\l]=i\left(\y_{\m\l}\hp_\n-\y_{\l\n}\hp_\m\right),\quad[J_\mn,\hx_\l]=i\left(\y_{\m\l}\hx_\n-\y_{\n\l}\hx_\m\right).
\ee
In particular, the components of the position operator $\hx_\m$ do not commute among themselves.
The \cc $\b$ has dimension of inverse mass square and may be identified with $1/M_P^2=L_P^2$, $M_P$ being the Planck mass.
Note that $\b$ can take both signs. If it is positive, the bound $p^2<1/\b$ must hold, and hence a maximal mass $1/\b$ is
predicted.

Snyder also showed that his algebra can be realized on canonical phase space of \coo $x$ and $p$, with Lorentz generators defined as
$J_\mn=x_\m p_\n-x_\n p_\m$.

In contrast with most common models of noncommutative geometry, the commutators are not linear in the phase
space variables: this allows them to be compatible with a linear action of the Lorentz symmetry, so that the \poi algebra is not deformed.
However, translations (generated by the $\hp_\m$) act in a nonlinear way on position variables.

The Snyder model can be interpreted as describing flat spacetime with a curved momentum space.
In fact, the subalgebra generated by $J_\mn$ and $\hx_\m$ is isomorphic to the de Sitter algebra $so(1,4)$, and hence
the Snyder momentum space has the same geometry as de Sitter spacetime.

\section{The Yang algebra}
Soon after the publication of Snyder's paper, Yang proposed a generalization where also the momentum variables do not commute,
like in de Sitter spacetime \cite{yan}.

The algebra is isomorphic to $so(1,5)$, with 15 generators,
$$[\hx_\m,\hx_\n]=i\b J_\mn,\quad[\hp_\m,\hp_\n]=i\a J_\mn,\quad[\hx_\m,\hp_\n]=i\y_\mn K,$$
$$[J_\mn,J_{\r\s}]=i\big(\y_{\m\r}J_{\n\s}-\y_{\m\s}J_{\n\r}+\y_{\n\r}J_{\m\s}-\y_{\n\s}J_{\m\r}\big),$$
$$[J_\mn,\hp_\l]=i\left(\y_{\m\l}\hp_\n-\y_{\l\n}\hp_\m\right),\quad[J_\mn,\hx_\l]=i\left(\y_{\m\l}\hx_\n-\y_{\n\l}\hx_\m\right),$$
\be[K,\hx_\m]=i\b \hp_\m,\quad[K,\hp_\m]=-i\a \hx_\m,\quad[J_\mn,K]=0.\ee
The deformation parameter $\a$ has dimension of inverse length square and may be identified with the cosmological constant,
while ${\b}$ is the same as in the Snyder model. Notice that both $\a$ and $\b$ can take positive or negative values, giving
rise to models with very different physical properties \cite{MM}. For example, positive $\a$ enforces the bound $\hx^2<1/\a^2$ and
analogously, for positive $\b$, $\hp^2<1/\b^2$.

The Yang algebra contains as subalgebras both the \des and the Snyder algebras, and therefore describes a noncommutative model
in a spacetime of constant curvature.
In order to close the algebra, Yang had to introduce a new generator $K$ which rotates positions into momenta, but whose physical
interpretation is not evident.

The previous algebra is invariant under a generalized Born duality \cite{bor,guo},
\be
\a\lra\b,\quad\hx_\m\to-\hp_\m,\quad\hp_\m\to\hx_\m,\quad J_\mn\lra J_\mn,\quad K\lra K.
\ee
The isomorphism with the $so(1,5)$ algebra can be obtained by identifying
\be\lb{iso}
M_\mn=J_\mn,\qquad M_{\mu4}=\hx_\m,\qquad M_{\mu5}=\hp_\m,\qquad M_{45}=K,
\ee
where $M_{AB}$ ($A,B=0,\dots,5$)  are the generators of $so(1,5)$.

\medskip

There also exists a different generalization of the Snyder algebra on curved space, known as triply special relativity (TSR),
that does not include $K$, but is nonlinear \cite{tsr}.

In particular, in that case the deformed Heisenberg subalgebra takes the form
\bd
[\hx_\m,\hx_\n]=i\b J_\mn,\quad[\hp_\m,\hp_\n]=i\a J_\mn,
\ed
\be[\hx_\m,\hp_\n]=i\Big(\y_\mn+\a\hx_\m\hx_\n+\b\hp_\m\hp_\n+\sqrt{\a\b}(\hx_\m\hp_\n+\hp_\m\hx_\n-J_\mn)\Big),\ee
and one can interpret the phase space as a coset space $SO(1,5)/SO(1,3)\times O(2)$.

Triply special relativity theory has been described as a deformation of the Galilei group in \cite{oko} and its Hamiltonian formulation
has been investigated in \cite{ban}.
A framework unifying it with the Yang model has been proposed in \cite{MS}.

\medskip

\noindent Two interpretations of the Yang algebra are possible:

\medskip
Take the algebra as it is, with its 15 generators \cite{luk1,MMM}. This allows one to construct the Hopf algebra structure, with the
related star product, twist, etc. exploiting the results known for generic orthogonal algebras \cite{esn2}
 However, in this case  one has to consider an extended phase space with scalar and tensorial degrees of freedom, whose interpretation
 is not obvious.

\medskip
Take a nonlinear realization on canonical phase space spanned by $x_\m$ and $p_\m$, with $J_\mn=x_\m p_\n-x_\n p_\n$ and $K=K(x,p)$ \cite{mel1},
in analogy with the representation given by Snyder for his model \cite{sny}.
In this case the interpretation is easier and one can include the Yang model in the same family of nonlinear realizations of $so(1,5)$ as TSR,
identifying the phase space with a coset space.
However, one can no longer define a Hopf algebra, star products etc.

\section{Realizations of the Yang-Poisson model}
We start following the second route, and discussing the classical limit of the Yang model, in which commutators are replaced by
Poisson brackets.
This makes the investigation much easier because of the absence of ordering problems. We call this limit Yang-Poisson model.

 We have \cite{mel3}
\bd
\{\hx_\m,\hx_\n\}=\b J_\mn,\quad\{\hp_\m,\hp_\n\}=\a J_\mn,\quad\{\hx_\m,\hp_\n\}=\y_\mn K,
\ed
\be\{K,\hx_\m\}=\b\hp_\m,\quad\{K,\hp_\m\}=-\a\hx_\m,
\ee
and look for an expression of $K(x,p)$ that satisfies the previous Poisson brackets.
\bd
\hx_\m=f(p^2,z)\,x_\m ,\qquad \hp_\m=g(x^2,z)\,p_\m,
\ed
\be
K=K(x^2,p^2,z),
\ee
where $z=x\bdot p$ and $f$ and $g$ are functions to be determined.

The only nontrivial brackets to be checked are those of the deformed Heisenberg algebra,
which give rise to partial differential equations. The equations derived from the $x$--$x$ and $p$--$p$ brackets
have solutions
\be
f=\sqrt{1-\b p^2+\f_1(z)},\qquad g=\sqrt{1-\a x^2+\f_2(z)},
\ee
with arbitrary functions $\f_1$ and $\f_2$, while the $x$--$p$ brackets give
\be
\f_1\f_2+\f_1+\f_2=\a\b z^2,\qquad K=fg,
\ee
with solution depending on one parameter $c$
\be
\f_1(z)={\sqrt{1+4c(1-c)z^2}-1\over2(1-c)},\quad\f_2(z)={\sqrt{1+4c(1-c)z^2}-1\over2c}.
\ee
Then,
\be
\hx_\m=\sqrt{1-\b p^2+\f_1(z)}\ x_\m,\quad \hp_\m=\sqrt{1-\a x^2+\f_2(z)}\ p_\m,
\ee
and
\be
K=\sqrt{\left[1-\b p^2+\f_1(z)\right]\left[1-\a x^2+\f_2(z)\right]}.
\ee

A particularly interesting solution is obtained by assuming symmetry under the exchange of $x$ and $p$, as is natural in
view of the Born duality of the model. In this case, $\f_1=\f_2=\f$, \ie $c=\ha$, and we obtain
\be
\f=\sqrt{1+\a\b z^2}-1,
\ee
and then
\be\lb{real}
\hx_\m=\sqrt{\sqrt{1+\a\b z^2}-\b p^2}\ x_\m,\quad \hp_\m=\sqrt{\sqrt{1+\a\b z^2}-\a x^2}\ p_\m.
\ee
This gives an exact realization of the Yang-Poisson model, symmetric for $x\lra p$ and $\a\lra\b$.
One can also write $K$ in terms of the original variables, as
\be
K=\sqrt{{1-\a\hx^2-\b\hp^2\over2}\left(1+\sqrt{1-4\a\b{\hx^2\hp^2-(\hx\bdot\hp)^2\over(1-\a\hx^2-\b\hp^2)^2}}\right)}.
\ee

More general realizations can also be found \cite{mel3,MMM1}.

\section{Realizations of the quantum Yang model}

In the quantum case, finding a realization is more difficult, and no general closed form is known.
One has therefore to resort to a perturbative calculation in the coupling parameters $\a$ and $\b$.

A perturbative calculation up to fourth order was first performed in \cite{mel1},
More elaborated methods were used in \cite{MMM}, where also representations in extended phase space were investigated.

Realizations can be found order by order, making an ansatz that includes the most general Lorentz-covariant terms.
For example, at first order in $\a$ and $\b$ one can find an Hermitian realization setting
\bd
\hx_\m=x_\m+\big(a_1\sqrt{\a\b}x_\m\xp+a_2\b x_\m p^2+a_3\b p_\m\px+a_4\sqrt{\a\b}p_\m x^2+{\rm h.c.}\big),
\ed
\bd
\hp_\m=p_\m+\big(b_1\sqrt{\a\b}p_\m\px+b_2\a p_\m x^2+b_3\a x_\m\xp+b_4\sqrt{\a\b}x_\m p^2+{\rm h.c.}\big),
\ed
\be
K=1+\big(h_1\a x^2+h_2\sqrt{\a\b}\xp+h_3\b p^2+{\rm h.c.}\big),
\ee
and substituting in (\ref{1}), obtaining constraints between the free parameters.

Remarkably, the simplest solution of these constraints symmetric in $\hx$ and $\hp$, is given at second order by \cite{mel1}
$$\hx_\m=x_\m-{\b\over4}x_\m p^2-{\b^2\over16}x_\m p^4+{\a\b\over8}x_\m\xp\px+{\rm h.c.}$$
\be
\hp_\m=p_\m-{\a\over4}p_\m x^2-{\a^2\over16}p_\m x^4+{\a\b\over8}p_\m\px\xp+{\rm h.c.},
\ee
with
\be
K=1-\ha\Big(\a x^2+\b p^2\Big)-{1\over8}\Big(\a x^2-\b p^2\Big)^2+{\a\b\over2}\,\xp\px.
\ee

Of course, this is noting but the expansion of the Yang-Poisson result (\ref{real}) in powers of $\a$ and $\b$.
However, going to higher orders in $\a$ and $\b$ one finds corrections with respect to the classical solution.

\section{Star product for the Yang algebra}

One of the most useful frameworks for the investigation of noncommutative geometry is that of Hopf algebras \cite{hop}.
This formalism implies the definition of a coalgebraic structure including a coproduct and an antipode.

Such formalism can be applied also to the Yang model, provided one takes all its generators $M_{AB}$ as primary variables
\cite{MMM,MMM2},
and therefore use a realization on extended phase space \cite{MMM1}.
One can then use the isomorphism (\ref{iso}) of the Yang algebra with $so(1,5)$ and the general results of \cite{esn2}, where
the Hopf algebra related to general orthogonal groups was computed. Notice that to this end it is necessary to define ''momenta"
$s^{AB}$ canonically conjugated to the variables $M_{AB}$, whose physical interpretation is not obvious.

We shall not illustrate the details here, referring the reader to the original literature.
We only recall that due to the noncommutativity, the addition law of the momenta $s^{AB}$ is deformed \cite{hop}.
Using the Hopf algebra formalism, this deformation can be expressed by means of a star product, whose knowledge can be useful, for
example, in the construction of a quantum field theory.

In our case the star product for plane waves can be defined as
\be
e^{{i\over2}s^{AB}M_{AB}}\star e^{{i\over2}t^{CD}M_{CD}}=e^{{i\over2}\cD^{AB}(s,t)M_{AB}},
\ee
where $s_{AB}$ and $t_{AB}$  are antisymmetric tensors that describe the ''momenta" conjugated to the primary variables $M_{AB}$,
and $\cD^{AB}$ encodes the deformed addition law.

Using a Weyl realization of the algebra, the star product can be calculated perturbatively \cite{MMM2}.
It may be useful to explicitly write down the four-dimensional expression of $\cD^{AB}(s,t)$ at first order:
setting $\cD^\m=\cD^{\m4}$, $\bar\cD^\m=\cD^{\m5}$, $\cD=\cD^{45}$, one has
\bea
\cD^\mn(s,t)&=&s^\mn+t^\mn-{1\over2}\Big(s^{\m\l}t^\n_{\ \l}+\b s^\m t^\n+\a\bs^\m\bt^\n+\sqrt{\a\b}(s^\m\bt^\n+\bs^\m t^\n)\cr
&&-(\m\lra\n)\Big),\cr
\cD^\m(s,t)&=&s^\m+t^\m-{1\over2}\left(s^{\m\l}t_\l-t^{\m\l}s_\l+\sqrt{\a\b}(s^\m t-st^\m)+\a(\bs^\m t-s\bt^\m)\right),\cr
\bar\cD^\m(s,t)&=& \bs^\m+\bt^\m-{1\over2}\left(s^{\m\l}\bt_\l-\bs_\l t^{\m\l}-\sqrt{\a\b}(\bs^\m t-s\bt^\m)+\b(s^\m t-st^\m)\right),\cr
\cD(s,t)&=&s+t-{1\over2}\left(s^\l\bt_\l-\bs^\l t_\l\right),
\eea
where $s^\mn$, $s^\m=s^{\m4}$, $\bs^\m=s^{\m5}$, $s=s^{45}$ are the four-dimensional components of $s^{AB}$,
conjugated to $J_\mn$, $x_\m$, $p_\m$ and $K$, resp. It is clear from these expressions that the four-dimensional components of the
momenta mix in the star product.

\section{Generalizations}
The Yang algebra can be generalized in several ways.
The simplest one is to admit negative values of the deformation parameters $\a$ or $\b$, obtaining
models based on the algebras $so(2,4)$ or $so(3,3)$ \cite{MMM1}. These models have different physical properties from
those with positive $\a$ and $\b$ \cite{MM}.

Another simple generalization was given in \cite{lez}, where linear combinations of the generators $\hx$ and $\hp$
by a parameter $\g$ where introduced.
In \cite{MS,mel1} still more general definitions of the Yang model in extended phase space were proposed,
unifying it with TSR.
Also a supersymmetric extension of the Yang algebra was investigated in \cite{LW}.

Finally, the possibility to include $\k$-deformations of both position and momentum space with parameters
$a_\m$ and $b_\m$ into the model of ref.~\cite{lez} was studied in \cite{luk,MMM2}. The generalized algebra reads
\bea\lb{def}
[x_\m,x_\n]&=&i\left(\b J_\mn+a_\m x_\n-a_\n x_\m\right),\cr
[p_\m,p_\n]&=&i\left(\a J_\mn+b_\m p_\n-b_\m p_\n\right),\cr
[x_\m,p_\n]&=&i\left(\y_\mn  h+b_\m x_\n-a_\n p_\m+\g J_\mn\right),\cr
[J_\mn,x_\l]&=&i\left(\y_{\m\l}x_\n-\y_{\n\l}x_\m+a_\m J_{\l\n}-a_\n J_{\l\m}\right),\cr
[J_\mn,p_\l]&=&i\left(\y_{\m\l}p_\n-\y_{\n\l}p_\m+b_\m J_{\l\n}-b_\n J_{\l\m}\right),\cr
[J_\mn,K]&=&i\left(b_\n x_\m-b_\m x_\n-a_\n p_\m+a_\m p_\n\right),\cr
[K,x_\m]&=&i\left(\b p_\m-\g x_\m-a_\m K\right),\cr
[K,p_\n]&=&i\left(-\a x_\m+\g p_\m+b_\m K\right).
\eea
This algebra is still isomorphic to $so(1,5)$, but now the action of the Lorentz group is deformed.
It is however still possible to obtain the coalgebra associated to it \cite{MMM2}.

\section{Applications}

A physical consequence of the Yang model is a deformation of the Heisenberg uncertainty relations \cite{MM}.
In fact, in its nonrelativistic 3-dimensional limit,
\be
\D x_i\D p_j\ge\ha\big|\langle[x_i,p_j]\rangle\big|=\ha\big|\langle K\rangle\big|\d_{ij}.
\ee
Clearly, in a phase space realization with $K=K(x,p)$ this gives rise to a generalized uncertainty principle,
similar to the one valid for TSR \cite{mig}.

For example, let us consider a one-dimensional toy model: at leading order in $\hbar$, $\a$ and $\b$,  one can use the realization (\ref{real}),
obtaining\footnote{In this section, we reintroduce explicitly $\hbar$.}
\be
\D\hx\D\hp\ge{\hbar\over2}\sqrt{1-\a(\D\hx)^2-\b(\D\hp^2)}.
\ee

It follows that
\be
(\D\hx)\ge\hbar\sqrt{{1-\b(\D\hp)^2\over\hbar^2\a+4(\D\hp)^2}},\qquad(\D\hp)\ge\hbar\sqrt{{1-\a(\D\hx)^2\over\hbar^2\b+4(\D\hx)^2}}.
\ee
Clearly, the uncertainty relations will strongly depend on the sign of the coupling constants.
For example, for $\a,\b>0$, the uncertainties satisfy the bounds $0\le\D p\le 1/\sqrt\b$, and $0\le\D x\le 1/\sqrt\a$, as is obvious because of
the limited range of variation of $x$ and $p$ in this case.
For $\a,\b<0$, instead, one gets the interesting relations $\D p\ge\hbar\sqrt{|\a|}/2$, and $\D x\ge\hbar\sqrt{|\b|}/2$.
A more complicated behaviour occurs if $\a$ and $\b$ have different sign \cite{MM}.

One may also calculate corrections to the dynamics of simple models due to the nontrivial symplectic structure,
with possible applications to astrophysical observations, again in analogy with the results of \cite{mig} for TSR.

Finally, a more ambitious goal would be to build a quantum field theory based on this framework. This would presumably require
the knowledge of the star product and hence the use of an extended phase space. In this case the physical interpretation
of the additional coordinates needs to be clarified.

\section{Final remarks}
We have reviewed the main properties of a model of noncommutative geometry on a curved background proposed by C.N. Yang in the fourties,
and have reported some developments and generalizations obtained in the last years.

The physical relevance of these models is that they somehow relate the description of spacetime at extremely small scale and at cosmological
scales, and that they may be connected to a low-energy limit of quantum gravity \cite{tsr}.
Some applications have been investigated, however much remains to be understood concerning the physical interpretation and the phenomenological
predictions of these models.

\end{document}